\documentclass[10pt,a4paper]{article}
\usepackage{amsfonts}
\usepackage{amssymb}
\usepackage{multicol}
\usepackage{graphicx}
\usepackage{float}
\usepackage{caption}
\usepackage{xcolor}
\usepackage[utf8]{inputenc}
\usepackage{amsmath}
\usepackage{authblk}
\usepackage[colorlinks=true,
            linkcolor=black,
            citecolor=black,
            urlcolor=black]{hyperref}
\usepackage{float}
\usepackage{subcaption}
\usepackage{url}
\usepackage{titlesec}
\usepackage{nameref}
\usepackage{epsfig}    
\usepackage{cleveref}
\usepackage{authblk}

\makeatletter \@addtoreset{equation}{section}
\renewcommand{\theequation}{\thesection.\arabic{equation}}
\def \be{\begin{equation}}
\def \ee{\end{equation}}
\def \bea{\begin{eqnarray}}
\def \eea{\end{eqnarray}}

\newcommand{\nc}{\newcommand}
\nc{\al}{\alpha} \nc{\bib}{\bibitem} \nc{\la}{\lambda}
\nc{\C}{\mbox{\hspace{1.24mm}\rule{0.2mm}{2.5mm}\hspace{-2.7mm} C}}
\nc{\R}{\mbox{\hspace{.04mm}\rule{0.2mm}{2.8mm}\hspace{-1.5mm} R}}
\input{tcilatex}
\begin{document}

\title{
\textbf{ Multifield quintessential inflation and Swampland constraints in the IDM}}

\renewcommand\Authands{, }
\author[1,3]{S. Saoud%
\thanks{\href{mailto:soulaimane_saoud@um5.ac.ma}{soulaimane\_saoud@um5.ac.ma}}}

\author[1,3]{M. A. Rbah%
\thanks{\href{mailto:Mohamedamin_rbah@um5.ac.ma}{Mohamedamin\_rbah@um5.ac.ma}}}

\author[1,3]{R. Sammani%
\thanks{\href{mailto:rajae_sammani@um5.ac.ma}{rajae\_sammani@um5.ac.ma}}}

\author[1,2,3]{E. H. Saidi%
\thanks{\href{mailto:e.saidi@um5r.ac.ma}{e.saidi@um5r.ac.ma}}}

\author[1,3]{R. Ahl Laamara%
\thanks{\href{mailto:r.ahllaamara@um5r.ac.ma}{r.ahllaamara@um5r.ac.ma}}}

\affil[1]{\small LPHE-MS, Faculty of Sciences, Mohammed V University, Rabat, Morocco}

\affil[2]{\small Hassan II Academy of Science and Technology, Kingdom of Morocco}

\affil[3]{\small Centre of Physics and Mathematics, CPM-Morocco}
\maketitle

\begin{abstract}
We investigate quintessential inflation in the Inert Doublet Model
coupled to a light quintessence field through an exponential
interaction. In the Einstein frame, the model forms a noncanonical
two-field system with a curved field-space geometry and a
Starobinsky-like inflationary plateau. We solve the covariant
background and perturbation equations and perform a parameter scan of
the exponential slope and the number of inflationary e-folds. The
scalar amplitude fixes the inflationary normalization independently at
each scan point, rather than selecting a particular value of the
quintessence coupling. We find that cumulative
entropy-to-curvature transfer significantly modifies the scalar
spectrum, while the tensor-to-scalar ratio remains compatible with
current CMB limits. The scalar spectral index provides the strongest
sensitivity to the exponential coupling. Covariant Swampland
diagnostics nevertheless reveal a persistent tension with a strong
order-unity de Sitter bound.
\end{abstract}

\section{Introduction}
\label{sec:introduction}

The accelerated expansion of the Universe, both during the primordial
inflationary era and at the present dark-energy-dominated epoch, remains
one of the central challenges of modern cosmology. Inflation provides a
compelling mechanism for generating the observed large-scale
homogeneity and isotropy of the Universe while promoting quantum
fluctuations to cosmological scales, thereby providing the seeds for
the formation of large-scale structure
\cite{guth1981inflationary}. Precision
measurements of the cosmic microwave background (CMB), including
temperature and polarization observations from \emph{WMAP},
\emph{Planck}, and \emph{BICEP/Keck}, strongly constrain the amplitude
and scale dependence of primordial fluctuations and favor inflationary
models characterized by relatively flat or plateau-like potentials
\cite{komatsu4538seven,ade2016planck,akrami2020planck}. A well-known
realization of this idea is Higgs inflation, in which a scalar field
non-minimally coupled to gravity develops a nearly flat potential in
the Einstein frame
\cite{salopek1989designing,bezrukov2008standard,kaiser1995primordial,
Kallosh_2014}. Nevertheless, the extrapolation of the Standard Model
scalar sector to very high energies raises questions related to vacuum
stability and motivates extensions containing additional scalar
degrees of freedom \cite{sher1989electroweak}.

At late times, the observed accelerated expansion is successfully
described within the standard $\Lambda$CDM framework
\cite{perlmutter1999measurements,akrami2020planck}. Despite its
phenomenological success, the extremely small observed value of the
vacuum-energy density motivates the consideration of dynamical
alternatives. Quintessence provides one such possibility, in which a
light scalar field slowly evolves along a sufficiently shallow
potential and generates an effective negative pressure at late times
\cite{Copeland:1997et,tsujikawa2013quintessence}. This naturally raises
the question of whether primordial inflation and present-day cosmic
acceleration can emerge within a common scalar-field framework rather
than being treated as completely unrelated phenomena.

A further theoretical motivation comes from attempts to embed
cosmological effective field theories within a consistent theory of
quantum gravity. The Swampland program aims to characterize those
low-energy effective theories that may admit an ultraviolet completion
and to distinguish them from theories that remain incompatible with
quantum-gravity expectations
\cite{brennan2017string,10.21468/SciPostPhys.18.6.173,Sammani:2025ydx}.
Of particular relevance for inflation and dark energy are the de
Sitter and refined de Sitter conjectures, which constrain either the
covariant gradient of the scalar potential or the smallest eigenvalue
of its covariant Hessian \cite{Ooguri_2019}. Schematically, the
gradient condition requires
\begin{equation}
 M_P\,\frac{\sqrt{G^{IJ}V_{,I}V_{,J}}}{V}
 \gtrsim {\cal O}(1),
 \label{eq:intro-ds-bound}
\end{equation}
where $G_{IJ}$ denotes the field-space metric. Such a requirement is
difficult to reconcile with the flat potentials generally preferred
by slow-roll inflation and dynamical dark-energy scenarios, and has
therefore motivated extensive investigations of multifield dynamics,
noncanonical field-space geometries, and alternative realizations of
accelerated expansion
\cite{heisenberg2018dark,akrami2019landscape,colgain2019testing}.

In this context, the Inert Doublet Model (IDM) provides a particularly
interesting extension of the Standard Model scalar sector. It contains
a second Higgs doublet protected by a discrete $\mathbb{Z}_2$
symmetry, allowing the lightest inert state to be stable and hence to
serve as a dark-matter candidate while avoiding tree-level
flavor-changing neutral currents \cite{justino2024constraints}. The IDM
has also been explored as an inflationary framework when the scalar
sector is non-minimally coupled to gravity, since the conformal
transformation to the Einstein frame naturally generates a
Starobinsky-like plateau at large field values \cite{Choubey_2017}.
These properties make the IDM a useful setting in which inflation,
dark matter, dark energy, and quantum-gravity consistency conditions
can be investigated within a common framework.

Motivated by these considerations, we study a quintessential
inflationary realization of the IDM in which the inert scalar sector
is coupled to a light quintessence field through an exponential
factor. After transforming from the Jordan frame to the Einstein
frame, the resulting theory is not equivalent to a canonical
single-field model but instead defines a genuinely noncanonical
two-field system evolving on a curved field-space manifold. We
therefore treat both the background and the perturbations
covariantly. In particular, we construct the adiabatic and entropy
directions, evaluate the turn rate of the background trajectory, and
follow the entropy-to-curvature conversion instead of assuming from
the outset that the quintessence direction behaves as a negligible
spectator.

An important consequence of this treatment concerns the normalization
of the primordial spectrum. The measured scalar amplitude does not
directly impose a bound on the quintessence coupling $\beta$.
Instead, it primarily fixes the overall inflationary normalization,
which in the large non-minimal-coupling regime is controlled by the
combination $\lambda_2/\xi_2^2$. The quintessence coupling affects the
background trajectory, the field-space dynamics, and the conversion
between entropy and curvature fluctuations, and must therefore be
constrained through the complete multifield evolution rather than
through a single-field normalization argument. For a representative
benchmark with $\beta=0.004$, $\xi_2=4.9\times10^4$, and
$N_*=60$, the normalized model yields a scalar spectral index close to
$n_s\simeq0.967$ and a tensor-to-scalar ratio
$r\simeq3.0\times10^{-3}$, consistent with current CMB constraints.

The multifield character of the model is also found to be
phenomenologically relevant. The inflationary trajectory exhibits no
sharp turn; rather, it undergoes a slow but cumulative bending over
several tens of e-folds. Although the instantaneous turn rate remains
small, the accumulated change of direction is non-negligible and is
accompanied by a sizeable entropy-to-curvature transfer. Consequently,
the final scalar spectrum differs appreciably from the result obtained
by applying the single-field relation $r=16\epsilon_H$. This provides
a concrete example in which a relatively mild departure from a
geodesic trajectory can nevertheless have an important impact on the
observable perturbation spectrum.

We furthermore reassess the Swampland implications using the full
covariant field-space quantities. In contrast with the identification
of the de Sitter slope parameter with the exponential coupling
$\beta$, the relevant quantity during inflation is the norm of the
complete multifield potential gradient. For the representative
benchmark, this covariant gradient remains of order a few
$10^{-2}$ throughout the CMB window, while the most negative
eigenvalue of the normalized covariant Hessian is also parametrically
smaller than unity. The model therefore does not realize a rapid-turn
mechanism capable of supporting inflation on an order-unity potential
gradient. Multifield effects substantially modify the inflationary
phenomenology, but they do not remove the quantitative tension with a
strong interpretation of the de Sitter Swampland conjecture in which
the relevant coefficients are assumed to be of order unity.

The late-time evolution provides a complementary effective
description of the same framework. After stabilization of the radial
inflationary direction, the residual quintessence sector is described
by a shallow exponential potential. Initializing the field near rest
during radiation domination and normalizing the residual potential
to the observed dark-energy density lead to an expansion history
extremely close to that of a cosmological constant. This result
demonstrates the compatibility of the late-time exponential sector
with the observed background evolution, but does not provide a
continuous dynamical matching from the end of inflation through
reheating.

The purpose of the present work is therefore not to claim that the
minimal construction completely resolves the Swampland tension, but
rather to determine quantitatively how the tension is modified once
the full curved-field-space and multifield dynamics are taken into
account. We show that the entropy sector and the bending of the
trajectory are essential for reliable predictions of the primordial
observables, whereas the strong order-unity de Sitter criterion remains
difficult to satisfy during the observationally relevant stage of
inflation. This distinction between multifield phenomenology and
Swampland consistency constitutes one of the central results of our
analysis.

The remainder of this paper is organized as follows.
Section~\ref{sec:swampland} introduces the IDM scalar sector coupled
to quintessence and formulates the corresponding covariant de Sitter
Swampland criteria. Section~\ref{sec:framework} presents the
inflationary reduction of the model and derives the Einstein-frame
potential and field-space geometry. Section~\ref{sec:dynamics}
develops the exact covariant background and perturbation dynamics,
including the adiabatic--entropy decomposition, the normalization of
the primordial spectrum, the CMB parameter scan, and the covariant
Swampland diagnostics. It also examines the effective late-time
quintessence evolution and the consistency of the reduced radial
trajectory. Finally, Section~\ref{sec:conclusion} summarizes the main
results and discusses the limitations and possible extensions of the
minimal scenario.

\section{dS conjecture and the IDM with quintessence}\label{sec:swampland}
In this section, we investigate the behavior of the Higgs potential of the Inert Doublet Model (IDM) in the presence of a quintessence field. In the first subsection, we revisit the scalar potential of the IDM and formulate its coupling to quintessence. In the second subsection, we examine the compatibility of the resulting potential with the refined de Sitter conjecture.
\subsection{The IDM scalar sector coupled to quintessence}
\label{subsec:IDM_quintessence}

The Inert Doublet Model (IDM) constitutes a minimal extension of the
Standard Model scalar sector obtained by introducing a second
$SU(2)_L$ scalar doublet, $H_2$, which is odd under an exact discrete
$\mathbb{Z}_2$ symmetry, whereas all Standard Model fields, including
the active Higgs doublet $H_1$, are even. The $\mathbb{Z}_2$ symmetry
forbids Yukawa interactions involving $H_2$ and prevents the inert
doublet from acquiring a vacuum expectation value at the electroweak
minimum. Consequently, the lightest neutral $\mathbb{Z}_2$-odd state
is stable and can provide a viable dark-matter candidate
\cite{branco2012theory,Treesukrat_2024}.

Assuming CP conservation and real scalar couplings, the IDM potential
is written as
\begin{align}
V_{\rm IDM}(H_1,H_2)
={}&
m_{11}^2 H_1^\dagger H_1
+m_{22}^2 H_2^\dagger H_2
+\lambda_1(H_1^\dagger H_1)^2
+\lambda_2(H_2^\dagger H_2)^2
\nonumber\\
&+\lambda_3(H_1^\dagger H_1)(H_2^\dagger H_2)
+\lambda_4|H_1^\dagger H_2|^2
+\frac{\lambda_5}{2}
\left[
(H_1^\dagger H_2)^2+\mathrm{h.c.}
\right].
\label{eq:IDM-potential}
\end{align}
After electroweak symmetry breaking, the two scalar doublets may be
parametrized as
\begin{equation}
H_1=
\begin{pmatrix}
0\\[1mm]
\dfrac{v+h}{\sqrt{2}}
\end{pmatrix},
\qquad
H_2=
\begin{pmatrix}
H^+\\[1mm]
\dfrac{S+iA}{\sqrt{2}}
\end{pmatrix},
\label{eq:IDM-doublets}
\end{equation}
where $v\simeq246~\mathrm{GeV}$ is the electroweak vacuum expectation
value, $h$ denotes the Standard Model--like Higgs boson, and
$H^\pm$, $S$, and $A$ are respectively the charged, neutral CP-even,
and neutral CP-odd inert scalars.

With the normalization adopted in Eq.~\eqref{eq:IDM-potential}, the
tree-level scalar masses are
\begin{align}
m_h^2 &= 2\lambda_1v^2,
&
m_{H^\pm}^2
&=
m_{22}^2+\frac{\lambda_3}{2}v^2,
\nonumber\\
m_S^2
&=
m_{22}^2
+\frac{\lambda_3+\lambda_4+\lambda_5}{2}v^2,
&
m_A^2
&=
m_{22}^2
+\frac{\lambda_3+\lambda_4-\lambda_5}{2}v^2.
\label{eq:IDM-masses}
\end{align}
It is convenient to introduce the Higgs-portal combination
\begin{equation}
\lambda_L
\equiv
\frac{\lambda_3+\lambda_4+\lambda_5}{2},
\label{eq:lambdaL}
\end{equation}
which controls the interaction between the Standard Model--like Higgs
boson and the neutral CP-even inert state $S$. Depending on the mass
ordering, either $S$ or $A$ can constitute the lightest inert particle
and therefore the dark-matter candidate.

Although the inert doublet has no Yukawa couplings, it retains the
standard electroweak gauge interactions through its covariant kinetic
term. Consequently, inert scalars can be pair-produced through
electroweak processes such as Drell--Yan production and vector-boson
fusion, while Higgs-mediated processes may contribute through the
scalar portal when kinematically accessible
\cite{Kalinowski_2020,LopezHonorez:2006,DeshpandeMa:1978}.
Phenomenologically, these processes lead to characteristic
missing-energy signatures accompanied by leptons or jets, and the
allowed parameter space is constrained by collider searches,
electroweak precision measurements, dark-matter direct detection, and
relic-density considerations
\cite{Dolle:2010,Goudelis_2013,Belanger:2015,Ghosh:2022}.
In the present work, these IDM properties provide the particle-physics
basis for embedding the inert scalar sector into a cosmological
framework containing both inflation and dynamical dark energy.

We introduce a light quintessence field $\phi$ through a
dilaton-inspired multiplicative coupling,
\begin{equation}
V(H_1,H_2,\phi)
=
e^{-\beta\phi/M_P}
\left[
V_{\rm IDM}(H_1,H_2)+\Lambda
\right],
\label{eq:IDM-quintessence-potential}
\end{equation}
where $\beta$ is dimensionless and $\Lambda$ denotes the residual
vacuum-energy scale relevant after stabilization of the Higgs
directions. Such exponential couplings arise naturally in effective
descriptions inspired by dilaton and quintessence constructions
\cite{Denef_2018,PhysRevD.62.043511,Usman:2015ida}.

The multiplicative structure of
Eq.~\eqref{eq:IDM-quintessence-potential} plays two conceptually
distinct roles in the cosmological evolution. At large field values,
it modulates the inflationary potential and participates in the full
multifield dynamics together with the non-minimally coupled inert
scalar direction. At late times, after the radial Higgs direction has
relaxed to its low-energy configuration, the surviving quintessence
sector is effectively described by a shallow exponential potential.
For a canonical exponential potential, a scalar-field-dominated
attractor gives accelerated expansion for sufficiently small slope,
with the familiar condition $\beta^2<2$
\cite{liddle1989power,kallosh2003dark}. In the complete model studied
here, however, the early-time dynamics cannot be reduced to this
single-field limit because the scalar fields evolve on a curved
field-space manifold.

This distinction is particularly important for the normalization of
the primordial spectrum. The observed scalar amplitude does not impose
a direct constraint on $\beta$. Instead, in the inflationary regime it
mainly fixes the overall height of the Einstein-frame potential,
controlled by the combination of the quartic and non-minimal
couplings, while $\beta$ affects the background trajectory and the
coupled adiabatic--entropy perturbations. We therefore determine the
inflationary predictions from the full covariant two-field evolution
rather than treating the quintessence direction as an effectively
decoupled spectator.

\subsection{Covariant de Sitter constraints on the coupled scalar sector}
\label{subsec:dS_conjecture}

The de Sitter Swampland conjecture was proposed as a criterion for
distinguishing low-energy effective theories that may admit a
consistent ultraviolet completion including quantum gravity
\cite{obied2018sitterspaceswampland}. For a collection of scalar
fields $\varphi^I$ described by a field-space metric
$\mathcal{G}_{IJ}$ and a positive scalar potential $V$, the original
de Sitter conjecture requires
\begin{equation}
M_P
\frac{
\sqrt{
\mathcal{G}^{IJ}
V_{,I}V_{,J}
}
}{V}
\geq
c_{\rm dS},
\label{eq:ds-gradient-criterion}
\end{equation}
where $c_{\rm dS}$ is a positive constant conjectured to be of order
unity. The quantity entering this condition is the invariant norm of
the complete potential gradient and is therefore independent of the
choice of scalar-field coordinates.

For later convenience, we define the effective covariant slope
\begin{equation}
c_{\rm eff}
\equiv
M_P
\frac{
\sqrt{
\mathcal{G}^{IJ}
V_{,I}V_{,J}
}
}{V}.
\label{eq:ceff-definition}
\end{equation}
The original conjecture can then be expressed as
\begin{equation}
c_{\rm eff}\geq c_{\rm dS}.
\label{eq:ceff-bound}
\end{equation}
The constant $c_{\rm dS}$ characterizes the conjectured quantum-gravity
bound and should not be confused with any coupling appearing in the
scalar potential.

A refined form of the conjecture allows a theory with a small
potential gradient provided that the potential possesses a sufficiently
tachyonic direction
\cite{Ooguri_2019,Garg:2018reu}. To formulate this second condition,
we introduce the mixed-index normalized covariant Hessian
\begin{equation}
\mathcal{H}^{I}{}_{J}
\equiv
\frac{M_P^2}{V}
\mathcal{G}^{IK}
\nabla_K\nabla_J V,
\label{eq:normalized-covariant-Hessian}
\end{equation}
where $\nabla_I$ is the Levi--Civita covariant derivative constructed
from $\mathcal{G}_{IJ}$. Denoting the smallest eigenvalue of
$\mathcal{H}^{I}{}_{J}$ by
\begin{equation}
\eta_{\min}
\equiv
\lambda_{\min}
\left(
\mathcal{H}^{I}{}_{J}
\right),
\label{eq:eta-min-definition}
\end{equation}
the refined de Sitter criterion requires that at least one of the
following conditions be satisfied:
\begin{equation}
c_{\rm eff}\geq c_{\rm dS},
\qquad\text{or}\qquad
\eta_{\min}\leq-c'_{\rm dS},
\label{eq:refined-dS-criterion}
\end{equation}
where $c_{\rm dS}$ and $c'_{\rm dS}$ are positive constants usually
assumed to be of order unity.

We now consider the coupled IDM potential introduced in
Eq.~\eqref{eq:IDM-quintessence-potential}; its logarithmic derivative along the quintessence direction is
\begin{equation}
M_P
\frac{V_{,\phi}}{V}
=
-\beta.
\label{eq:phi-logarithmic-gradient}
\end{equation}
Equation~\eqref{eq:phi-logarithmic-gradient} is an exact consequence
of the multiplicative exponential coupling. It determines the
coordinate derivative of the potential along the $\phi$ direction,
but it does not determine the norm of the complete gradient in
Eq.~\eqref{eq:ceff-definition}.

Indeed, the full scalar sector also contains the components of the
two Higgs doublets. Therefore,
\begin{equation}
\mathcal{G}^{IJ}V_{,I}V_{,J}
=
\mathcal{G}^{\phi\phi}V_{,\phi}^2
+
2\mathcal{G}^{\phi a}V_{,\phi}V_{,a}
+
\mathcal{G}^{ab}V_{,a}V_{,b},
\label{eq:gradient-decomposition-general}
\end{equation}
where the indices $a,b$ denote the scalar directions belonging to
the IDM sector. Introducing $v\equiv\ln V$, so that
$v_{,\phi}=-\beta/M_P$ and $v_{,a}=V_{,a}/V$,
Eq.~\eqref{eq:gradient-decomposition-general} gives
\begin{equation}
c_{\rm eff}^{\,2}
=
\beta^2\mathcal{G}^{\phi\phi}
-2\beta M_P\mathcal{G}^{\phi a}v_{,a}
+M_P^2\mathcal{G}^{ab}v_{,a}v_{,b}.
\label{eq:ceff-general-decomposition}
\end{equation}
Therefore, the identification $c_{\rm eff}=|\beta|$ holds if and
only if
\begin{equation}
\beta^2\left(\mathcal{G}^{\phi\phi}-1\right)
-2\beta M_P\mathcal{G}^{\phi a}v_{,a}
+M_P^2\mathcal{G}^{ab}v_{,a}v_{,b}
=0.
\label{eq:beta-identification-condition}
\end{equation}
For a canonically normalized quintessence direction,
$\mathcal{G}^{\phi\phi}=1$, and for $\beta\neq0$, this condition is
equivalently written as
\begin{equation}
\mathcal{G}^{\phi a}v_{,a}
=
\frac{M_P}{2\beta}\,
\mathcal{G}^{ab}v_{,a}v_{,b}.
\label{eq:beta-identification-canonical}
\end{equation}
The conditions
$\mathcal{G}^{\phi\phi}=1$,
$\mathcal{G}^{\phi a}=0$, and
$v_{,a}=0$ constitute a sufficient realization of
Eq.~\eqref{eq:beta-identification-condition}, since both sides of
Eq.~\eqref{eq:beta-identification-canonical} then vanish identically.
In a generic multifield theory with a non-diagonal field-space metric,
these conditions are sufficient but not mathematically necessary,
because contributions involving the mixed inverse-metric components
and the gradients along the additional scalar directions may combine
to satisfy the same equality
\cite{Andriot:2025multifield,Payeur:2024curved}.
For the present model, however, the field-space metric is block
diagonal, $\mathcal{G}^{\phi a}=0$, and no such cancellation occurs.
In the canonically normalized limit, the positive definiteness of
$\mathcal{G}^{ab}$ then implies $v_{,a}=0$.

For the multiplicative potential introduced in
Eq.~\eqref{eq:IDM-quintessence-potential}, we define
\begin{equation}
W(H_1,H_2)\equiv V_{\rm IDM}(H_1,H_2)+\Lambda.
\end{equation}
The logarithm of the complete potential and its derivatives along the
IDM directions are therefore given by
\begin{equation}
v\equiv\ln V
=-\frac{\beta\phi}{M_P}+\ln W,
\qquad
v_{,a}
=\partial_a\ln W
=\frac{W_{,a}}{W}.
\label{eq:logarithmic-IDM-gradient}
\end{equation}
At a stationary configuration of the IDM sector, $W_{,a}=0$ and,
provided that $W\neq0$, the logarithmic gradient along the IDM
directions vanishes. This statement refers exclusively to the IDM
directions and does not imply that the complete potential is constant,
since its derivative along the quintessence direction remains
\begin{equation}
v_{,\phi}=-\frac{\beta}{M_P},
\end{equation}
which is nonzero for $\beta\neq0$.

During inflation, however, the relevant IDM direction is dynamical
and is not generally located at a stationary configuration. Its
contribution to the covariant gradient must therefore be retained.
Moreover, the nonminimal gravitational coupling induces a nontrivial
field-space geometry for which $\mathcal{G}^{\phi\phi}$ is not
generally equal to unity. Consequently, the complete covariant slope
cannot, in general, be identified with $|\beta|$ during inflation.
The explicit reduced field-space metric and the corresponding
model-dependent expression for $c_{\rm eff}$ are derived below.

After restricting the inflationary background to the effective
two-dimensional scalar manifold constructed below, the tangent and
normal unit vectors form a complete orthonormal basis on that reduced
manifold. Only within this two-field description can the covariant
gradient be decomposed into adiabatic and entropic components. Let
$e_\sigma^I$ denote the unit vector tangent to the background
trajectory and $e_s^I$ the orthogonal unit vector. The corresponding
potential derivatives are defined by
\begin{equation}
V_{,\sigma}
\equiv e_\sigma^I V_{,I},
\qquad
V_{,s}
\equiv e_s^I V_{,I}.
\label{eq:adiabatic-entropic-gradient}
\end{equation}
Consequently, the norm of the potential gradient on the reduced
two-dimensional manifold becomes
\begin{equation}
\mathcal{G}^{IJ}V_{,I}V_{,J}
=
V_{,\sigma}^{2}+V_{,s}^{2}.
\label{eq:trajectory-gradient-decomposition}
\end{equation}
A nonzero entropic component $V_{,s}$ produces a bending of the
background trajectory. Such a curved trajectory may lead to
entropy-to-curvature transfer and can modify the relation between
the potential gradient and the inflationary observables
\cite{AchucarroPalma2019}.

After stabilization at a stationary IDM configuration,
$W_{,a}=0$ and therefore $v_{,a}=0$. The general expression for the
covariant slope then reduces to
\begin{equation}
c_{\rm eff}^{\rm late}
=
|\beta|
\sqrt{\mathcal{G}^{\phi\phi}_{\rm late}}.
\label{eq:late-time-ceff-general}
\end{equation}
If the remaining quintessence direction becomes canonically normalized,
$\mathcal{G}^{\phi\phi}_{\rm late}\simeq1$, one obtains
\begin{equation}
c_{\rm eff}^{\rm late}
\simeq|\beta|.
\label{eq:late-time-ceff}
\end{equation}
The identification of the exponential coupling with the de Sitter
slope is therefore valid only in the stabilized, effectively
single-field regime. It is not a general property of the
inflationary multifield system.

The distinction between the complete covariant gradient and the
exponential slope $\beta$ is therefore essential throughout the
cosmological evolution. During the multifield regime, the relevant
Swampland quantities depend on the full scalar potential, its
covariant derivatives, and the geometry of field space. In the
effectively single-field late-time regime, these quantities reduce to
their standard quintessence expressions once the IDM directions have
stabilized. The inflationary and late-time epochs must consequently be
examined within their respective dynamical regimes, without
identifying $\beta$ with the complete de Sitter slope during
inflation.

\section{Quintessential IDM in the Early Universe}
\label{sec:framework}

We formulate the inflationary sector of the quintessential Inert
Doublet Model in the Einstein frame, keeping the radial normalization
consistent with the doublet kinetic term. The reduced field-space
metric is retained without a large-field expansion. The quartic
truncation of the potential and the strong-coupling relations used
to display its plateau structure are stated separately.

\subsection{Jordan-frame formulation and inflationary field-space geometry}
\label{subsec:jordan-frame}
\label{subsec:radial-geometry}

The scalar--gravitational sector of the Jordan-frame action is
\begin{align}
S_J=\int d^4x\,\sqrt{-g}\,\Bigg[
&\frac{M_P^2}{2}R
+\xi_1H_1^\dagger H_1R+\xi_2H_2^\dagger H_2R
-\frac12g^{\mu\nu}\partial_\mu\phi\,\partial_\nu\phi
\nonumber\\
&-(D_\mu H_1)^\dagger(D^\mu H_1)
-(D_\mu H_2)^\dagger(D^\mu H_2)
-V(H_1,H_2,\phi)\Bigg],
\label{eq:Jordan-action}
\end{align}
where $M_P$ is the reduced Planck mass and the coupled potential is
defined in Eq.~\eqref{eq:IDM-quintessence-potential}. The constants
$\xi_1$ and $\xi_2$ are the non-minimal couplings of the active and
inert doublets, respectively.

For the homogeneous scalar background, the gauge fields are set to
zero. We introduce
\begin{equation}
F(H_1,H_2)\equiv\Omega^2
=1+\frac{2\xi_1H_1^\dagger H_1+2\xi_2H_2^\dagger H_2}{M_P^2},
\label{eq:conformal-factor-general}
\end{equation}
and work on the branch $F>0$. The Weyl transformation
\begin{equation}
\hat g_{\mu\nu}=F g_{\mu\nu}
\label{eq:Weyl-transformation}
\end{equation}
brings the gravitational sector into Einstein--Hilbert form
\cite{kaiser2010conformal,Tang_2021}.

Let $h_{IJ}$ denote the Jordan-frame scalar metric, defined by
$\mathcal L_{{\rm kin},J}
=-\tfrac12h_{IJ}g^{\mu\nu}\partial_\mu\Phi^I\partial_\nu\Phi^J$.
After the conformal transformation, the scalar action reads
\begin{equation}
S_E=\int d^4x\,\sqrt{-\hat g}\left[
\frac{M_P^2}{2}\hat R
-\frac12G_{IJ}(\Phi)\hat g^{\mu\nu}
\partial_\mu\Phi^I\partial_\nu\Phi^J
-V_E(\Phi)\right],
\label{eq:Einstein-action-general}
\end{equation}
with
\begin{equation}
V_E=\frac{V}{F^2},
\label{eq:Einstein-potential-general}
\end{equation}
and
\begin{equation}
G_{IJ}=\frac{h_{IJ}}{F}
+\frac{3M_P^2}{2F^2}F_{,I}F_{,J}.
\label{eq:field-space-general}
\end{equation}
Here $h_{IJ}=\delta_{IJ}$ in the original canonically normalized
real Cartesian fields. A change to radial coordinates must also
transform $h_{IJ}$. Since $F$ is independent of $\phi$, the
quintessence kinetic coefficient is $G_{\phi\phi}=\frac{1}{F},
G^{\phi\phi}=F.$


We restrict the homogeneous inflationary background to the branch
$\overline{H}_1=0$ and choose a fixed gauge orientation for the inert
doublet. This restriction concerns the background trajectory and does
not remove the orthogonal degrees of freedom from the complete theory.
Using the inert-doublet parametrization introduced previously, we
define the gauge-invariant radial variable by
\begin{equation}
H_2^\dagger H_2
\equiv \rho^2
=
H^+H^-+\frac{S^2+A^2}{2},
\qquad \rho\geq0.
\label{eq:rho-definition}
\end{equation}

A homogeneous doublet with nonvanishing norm can be brought, by an
electroweak gauge transformation, to a neutral and real
representative. We therefore choose
\begin{equation}
\overline{H}_2
=
\begin{pmatrix}
0\\
\rho
\end{pmatrix},
\qquad
\overline{S}=\sqrt{2}\,\rho,
\qquad
\overline{A}=\overline{H}^{\pm}=0.
\label{eq:neutral-radial-background}
\end{equation}
The bars emphasize that these relations apply only to the homogeneous
background values. In particular, setting
$\overline{A}=\overline{H}^{\pm}=0$ specifies the gauge orientation of
the inflationary trajectory and should not be interpreted as removing
these fields from the original IDM. The corresponding angular
directions are gauge directions along this background, whereas the
physical scalar fluctuations orthogonal to the reduced trajectory
must be treated separately when assessing its stability.

For vanishing background gauge fields, the inert-doublet kinetic term
evaluated on Eq.~\eqref{eq:neutral-radial-background} becomes
\begin{equation}
\mathcal L_{{\rm kin},J}^{(\rho)}
=
-(\partial_\mu\rho)(\partial^\mu\rho)
=
-\frac12\,2(\partial\rho)^2.
\label{eq:Jordan-radial-normalization}
\end{equation}
Comparing this result with the general Jordan-frame kinetic
convention introduced above shows that the radial component of the
reduced scalar metric is $h_{\rho\rho}=2$. Since the quintessence
field is canonically normalized in the Jordan frame and no mixed
kinetic term is generated at this stage, the metric restricted to the
coordinates $(\rho,\phi)$ is
\begin{equation}
h_{IJ}
=
\operatorname{diag}(2,1).
\label{eq:reduced-Jordan-metric}
\end{equation}

Substituting the background conditions
$\overline{H}_1=0$ and
$\overline{H}_2^\dagger\overline{H}_2=\rho^2$
into the general conformal factor of
Eq.~\eqref{eq:conformal-factor-general} gives, for $\xi_2>0$,
\begin{equation}
F(\rho)
=
1+\frac{2\xi_2\rho^2}{M_P^2}.
\label{eq:F-rho}
\end{equation}
The positivity condition $F>0$ is therefore automatically satisfied
along this trajectory for nonnegative $\xi_2$.
The restricted classical Jordan-frame potential is
\begin{equation}
V_J(\rho,\phi)
=e^{-\beta\phi/M_P}
\left(m_{22}^2\rho^2+\lambda_2\rho^4+\Lambda\right).
\label{eq:reduced-Jordan-potential-full}
\end{equation}
In the inflationary regime, where
$|m_{22}^2|\rho^2\ll\lambda_2\rho^4$ and
$|\Lambda|\ll\lambda_2\rho^4$, we use the quartic approximation
\begin{equation}
V_J\simeq V_J^{(4)},
\qquad
V_J^{(4)}\equiv\lambda_2\rho^4e^{-\beta\phi/M_P}.
\label{eq:large-field-Jordan-potential}
\end{equation}
The other IDM quartic couplings do not contribute to the background
potential at $H_1=0$, but remain relevant to transverse stability.

Using Eq.~\eqref{eq:field-space-general} with $h_{\rho\rho}=2$
gives the exact reduced radial kinetic coefficient
\begin{equation}
K(\rho)=\frac{2}{F}
+\frac{24\xi_2^2\rho^2}{M_P^2F^2}.
\label{eq:K-rho}
\end{equation}
Thus,
\begin{equation}
\mathcal L_{{\rm kin},E}
=-\frac12K(\rho)(\partial\rho)^2
-\frac{1}{2F(\rho)}(\partial\phi)^2,
\label{eq:two-field-kinetic}
\end{equation}
where contractions use the Einstein-frame spacetime metric.

An exactly canonical radial coordinate is defined, up to an
additive constant, by
\begin{equation}
\frac{d\chi_\rho}{d\rho}=\sqrt{K(\rho)},
\label{eq:canonical-radial-field}
\end{equation}
or equivalently,
\begin{equation}
\chi_\rho(\rho)-\chi_\rho(\rho_0)
=\int_{\rho_0}^\rho du\,
\sqrt{\frac{2}{F(u)}
+\frac{24\xi_2^2u^2}{M_P^2F(u)^2}}.
\label{eq:canonical-radial-integral}
\end{equation}
In particular, $K\to2$ and
$\chi_\rho(\rho)-\chi_\rho(0)\simeq\sqrt2\rho$ near the
origin. Radial canonical normalization leaves the quintessence
kinetic coefficient field dependent:
\begin{equation}
ds_{\rm fs}^2=d\chi_\rho^2
+\frac{d\phi^2}{F\bigl(\rho(\chi_\rho)\bigr)}.
\label{eq:canonical-field-space-exact}
\end{equation}

For later comparison with the stabilized regime, the reduced
field-space line element shows that
$\mathcal{G}_{\phi\phi}=F^{-1}$ and hence
$\mathcal{G}^{\phi\phi}=F$. For $\xi_1=0$, the inert vacuum satisfies
$\rho_{\rm late}=0$, so that $F(\rho_{\rm late})=F(0)=1$.
Consequently, $\mathcal{G}^{\phi\phi}_{\rm late}=1$, confirming the
late-time reduction $c_{\rm eff}^{\rm late}=|\beta|$ discussed in
Eq.~\eqref{eq:late-time-ceff}.

The conformal contribution to $K$ dominates when
\begin{equation}
\frac{24\xi_2^2\rho^2/(M_P^2F^2)}{2/F}
=6\xi_2\frac{F-1}{F}\gg1.
\label{eq:strong-coupling-condition}
\end{equation}
This condition is satisfied for $\xi_2\gg1$ and $F\gg1$.
Within this regime, integration of the dominant kinetic term and
a corresponding choice of the additive matching constant give
\begin{equation}
\chi_\rho\simeq\sqrt{\frac32}\,M_P\ln F,
\label{eq:chi-F}
\end{equation}
and hence
\begin{equation}
F\simeq\exp\left(\sqrt{\frac23}\frac{\chi_\rho}{M_P}\right).
\label{eq:F-chi}
\end{equation}
These relations are strong-coupling approximations and are not used
as exact changes of variables in the numerical evolution.

By contrast, the algebraic identities
\begin{equation}
\rho^2=\frac{M_P^2}{2\xi_2}(F-1),
\label{eq:rho2-F}
\end{equation}
and
\begin{equation}
\rho^4=\frac{M_P^4}{4\xi_2^2}(F-1)^2
\label{eq:rho4-F}
\end{equation}
follow exactly from Eq.~\eqref{eq:F-rho}. Combining them with
Eq.~\eqref{eq:F-chi} gives
\begin{equation}
\rho\simeq\frac{M_P}{\sqrt{2\xi_2}}
\left[e^{\sqrt{2/3}\,\chi_\rho/M_P}-1\right]^{1/2}.
\label{eq:rho-chi-strong-coupling}
\end{equation}
Only after imposing the additional large-field limit $F\gg1$ does
this reduce to
\begin{equation}
\rho\simeq\frac{M_P}{\sqrt{2\xi_2}}
\exp\left(\frac{\chi_\rho}{\sqrt6\,M_P}\right).
\label{3.10}
\end{equation}
The $(F-1)$ dependence must be retained when deriving the plateau
shape. Likewise, the exponential representation of the metric is
approximate:
\begin{equation}
ds_{\rm fs}^2\simeq d\chi_\rho^2
+e^{-\sqrt{2/3}\,\chi_\rho/M_P}d\phi^2.
\label{eq:canonical-field-space}
\end{equation}

\subsection{Einstein-frame effective potential}
\label{subsec:effective-potential}

Applying Eq.~\eqref{eq:Einstein-potential-general} to the restricted
classical potential gives
\begin{equation}
V_E=\frac{e^{-\beta\phi/M_P}}{F^2}
\left(m_{22}^2\rho^2+\lambda_2\rho^4+\Lambda\right).
\label{eq:VE-reduced-full}
\end{equation}
The inflationary approximation is $V_E\simeq V_E^{(4)}$, with
\begin{equation}
V_E^{(4)}(\rho,\phi)
\equiv\lambda_2\frac{\rho^4}{F^2}e^{-\beta\phi/M_P}.
\label{eq:VE-rho-phi}
\end{equation}
Equation~\eqref{eq:rho4-F} provides the exact rewriting of this
quartic potential,
\begin{equation}
V_E^{(4)}
=\frac{\lambda_2M_P^4}{4\xi_2^2}e^{-\beta\phi/M_P}
\left(1-\frac1F\right)^2.
\label{eq:VE-F}
\end{equation}
Using the approximate canonical relation in Eq.~\eqref{eq:F-chi}
then yields
\begin{equation}
V_E^{(4)}(\chi_\rho,\phi)
\simeq\frac{\lambda_2M_P^4}{4\xi_2^2}e^{-\beta\phi/M_P}
\left[1-\exp\left(-\sqrt{\frac23}
\frac{\chi_\rho}{M_P}\right)\right]^2.
\label{3.14}
\end{equation}
This expression displays the Starobinsky-like plateau in the
strong-coupling regime. Its height is controlled by
$\lambda_2/\xi_2^2$ for a specified quintessence trajectory.
The scalar amplitude fixes this inflationary normalization and
does not, by itself, establish an upper bound on $\beta$.
The nontrivial metric in Eq.~\eqref{eq:canonical-field-space-exact}
requires a covariant two-field calculation of the observables.

For numerical evolution, we use
\begin{equation}
r\equiv\frac{\rho}{M_P},\qquad
y\equiv\frac{\phi}{M_P},\qquad F(r)=1+2\xi_2r^2.
\label{eq:dimensionless-radial-fields}
\end{equation}
The exact reduced line element is
\begin{equation}
\begin{aligned}
ds_{\rm fs}^2&=M_P^2\left[A(r)dr^2+B(r)dy^2\right],\\
A(r)&=\frac2F+\frac{24\xi_2^2r^2}{F^2},
\qquad B(r)=\frac1F.
\end{aligned}
\label{eq:exact-reduced-metric}
\end{equation}
We define the dimensionless quartic potential by
\begin{equation}
U(r,y)\equiv\frac{V_E^{(4)}(M_Pr,M_Py)}{M_P^4}
=\lambda_2\frac{r^4}{F^2}e^{-\beta y}.
\label{eq:exact-reduced-potential}
\end{equation}
This formulation retains the full reduced metric while adopting the
quartic truncation of the IDM potential. Neither the logarithmic
canonical-field approximation nor a slow-roll reduction is required
for the background equations developed below.

\section{Covariant multifield dynamics and parameter scan}
\label{sec:dynamics}

We now investigate the inflationary dynamics of the reduced two-field
system without imposing a single-field slow-roll approximation. The
background trajectory, the two independent scalar mode functions, and
the tensor mode are evolved consistently on the same numerical
solution. This treatment preserves covariance under field
redefinitions and accounts for the nontrivial geometry of the scalar
manifold
\cite{Gordon:2000hv,GrootNibbelink:2001qt,Gong:2011uw,Dias:2015rca}.
The exponential slope $\beta$ is treated as a scan parameter, while the
observed scalar amplitude is used to determine the normalization of the
inflationary potential at each accepted point.

\subsection{Exact background system and numerical prescription}
\label{subsec:covariant-background}

Field-space indices refer initially to the dimensionful coordinates
$\Phi^I=(\rho,\phi)$. The numerical integration is subsequently
performed using the dimensionless variables defined in
Eq.~\eqref{eq:dimensionless-radial-fields}. Throughout the analysis, we
use the exact reduced metric of
Eq.~\eqref{eq:exact-reduced-metric} and the reduced potential of
Eq.~\eqref{eq:exact-reduced-potential}. In particular, the full radial
coefficient, including its $2/F$ contribution, is retained in both the
background and perturbation calculations. Such a curved field-space
structure is a generic consequence of transforming multifield models
with nonminimal gravitational couplings to the Einstein frame
\cite{Kaiser:2012ak,Kaiser:2013sna}.

The homogeneous scalar equations obtained by varying the
Einstein-frame action can be written in cosmic time as
$\mathcal{D}_t\dot{\Phi}^I+3H\dot{\Phi}^I+G^{IJ}V_{,J}=0$, where
$\mathcal{D}_t$ denotes the field-space covariant derivative along the
background trajectory. Introducing the number of e-folds,
$N=\ln a$, gives
\begin{equation*}
\dot{\Phi}^I=H\Phi^{I\prime},
\qquad
\mathcal{D}_t\dot{\Phi}^I
=
H^2\left(
\Phi^{I\prime\prime}
+\Gamma^I{}_{JK}\Phi^{J\prime}\Phi^{K\prime}
-\epsilon_H\Phi^{I\prime}
\right),
\end{equation*}
where a prime denotes differentiation with respect to $N$ and
$\epsilon_H=-\dot H/H^2$. Dividing the field equation by $H^2$ and
using the Friedmann constraint to eliminate $H^2$ gives
\begin{equation}
\Phi^{I\prime\prime}
+\Gamma^I{}_{JK}\Phi^{J\prime}\Phi^{K\prime}
+(3-\epsilon_H)
\left[
\Phi^{I\prime}
+M_P^2G^{IJ}\partial_J\ln V
\right]
=0.
\label{eq:background-covariant-section4}
\end{equation}
The quantities appearing in this equation satisfy
\begin{equation}
\epsilon_H
=
\frac{G_{IJ}\Phi^{I\prime}\Phi^{J\prime}}{2M_P^2},
\qquad
H^2
=
\frac{V}{M_P^2(3-\epsilon_H)}.
\label{eq:friedmann-section4}
\end{equation}
The first relation follows from the Raychaudhuri equation, whereas the
second is the exact Friedmann constraint expressed in terms of
$\epsilon_H$. No slow-roll approximation has been used in obtaining
Eqs.~\eqref{eq:background-covariant-section4} and
\eqref{eq:friedmann-section4}.

To obtain the equations used in the numerical integration, we write
$\varphi^i=(r,y)$ and factor the constant $M_P^2$ from the reduced
field-space line element. Equation~\eqref{eq:background-covariant-section4}
then takes the dimensionless form
\begin{equation*}
\varphi^{i\prime\prime}
+\Gamma^i{}_{jk}\varphi^{j\prime}\varphi^{k\prime}
+(3-\epsilon_H)
\left[
\varphi^{i\prime}
+g^{ij}\partial_j\ln U
\right]
=0.
\end{equation*}
Because the reduced metric is diagonal and depends only on $r$, the
only connection coefficients required for the homogeneous evolution,
together with the relevant logarithmic derivatives of the potential,
are
\begin{equation*}
\Gamma^r{}_{rr}=\frac{A_{,r}}{2A},
\qquad
\Gamma^r{}_{yy}=-\frac{B_{,r}}{2A},
\qquad
\Gamma^y{}_{ry}=\Gamma^y{}_{yr}
=\frac{B_{,r}}{2B},
\end{equation*}
and
\begin{equation*}
\partial_r\ln U=\frac{4}{rF},
\qquad
\partial_y\ln U=-\beta.
\end{equation*}
Substitution of these expressions into the radial and quintessence
components of the covariant equation yields
\begin{align}
r''
+\frac{A_{,r}}{2A}r'^2
-\frac{B_{,r}}{2A}y'^2
+(3-\epsilon_H)
\left(
r'+\frac{4}{ArF}
\right)
&=0,
\label{eq:r-background-section4}\\
y''
+\frac{B_{,r}}{B}r'y'
+(3-\epsilon_H)
\left(
y'-\beta F
\right)
&=0.
\label{eq:y-background-section4}
\end{align}
The terms proportional to $A_{,r}$ and $B_{,r}$ are the geometric
contributions generated by the Levi--Civita connection of the curved
field space. In the second equation, the factor $F$ multiplying
$\beta$ arises from the inverse metric component
$g^{yy}=B^{-1}=F$, rather than from a modification of the exponential
potential.

Finally, expressing the kinetic contraction in terms of the
dimensionless fields gives
\begin{equation}
\epsilon_H
=
\frac12\left(
Ar'^2+By'^2
\right).
\label{eq:epsilon-section4}
\end{equation}
Both field velocities therefore contribute to the departure from the
quasi-de Sitter regime. We define the end of inflation as the first
point along the numerical trajectory at which $\epsilon_H=1$.

We fix $\xi_2=4.9\times10^4$ and scan
\begin{equation}
0\leq\beta\leq0.020,
\qquad
N_*\in\{50,55,60\},
\label{eq:scan-domain}
\end{equation}
with spacing $\Delta\beta=0.002$. The background is initialized more than 16 e-folds before the earliest
pivot used in the scan, using the leading attractor velocities
\begin{equation}
r'=-\frac{4}{ArF},
\qquad
y'=\beta F.
\label{eq:initial-attractor-velocities}
\end{equation}
For each numerical trajectory, $N_{\rm end}$ is defined by the first
occurrence of $\epsilon_H=1$. The pivot time is then identified by
\begin{equation}
N_{\rm end}-N_{\rm pivot}=N_*,
\qquad
N_*\in\{50,55,60\}.
\label{eq:pivot-definition}
\end{equation}
Since only differences in the number of e-folds are physically
relevant, the origin of the plotted time coordinate is shifted so that
$N-N_{\rm pivot}=0$ at pivot crossing. The additive origin of $y$ is
similarly shifted so that $y_*=1$. This is a field-origin convention;
a shift in $y_*$ is degenerate with a rescaling of $\lambda_2$ in
Eq.~\eqref{eq:exact-reduced-potential}.

\subsection{Covariant scalar and tensor perturbations}
\label{subsec:adiabatic-entropy}

The scalar and tensor perturbations are evolved along the exact
numerical background trajectory. Throughout the calculation, we retain
the full field-space metric and the reduced potential without
introducing any large-field or single-field approximation.

The background velocity defines a preferred direction in field space.
Its norm and the corresponding unit tangent vector are
\begin{equation}
\dot{\sigma}^2
=
G_{IJ}\dot{\Phi}^I\dot{\Phi}^J,
\qquad
e_\sigma^I
=
\frac{\dot{\Phi}^I}{\dot{\sigma}}.
\label{eq:adiabatic-vector-section4}
\end{equation}
Thus, $e_\sigma^I$ identifies the instantaneous adiabatic direction,
namely the direction followed by the homogeneous trajectory. Since the
reduced system contains two scalar degrees of freedom, the orthogonal
direction is uniquely fixed up to an overall sign by
\begin{equation*}
G_{IJ}e_\sigma^I e_s^J=0,
\qquad
G_{IJ}e_s^I e_s^J=1.
\end{equation*}
Applying these conditions to the diagonal metric of
Eq.~\eqref{eq:exact-reduced-metric} gives
\begin{equation}
e_s^I
=
\frac{M_P}{\sigma'}
\left(
-\sqrt{\frac{B}{A}}\,y',
\sqrt{\frac{A}{B}}\,r'
\right),
\qquad
\sigma'
=
M_P\sqrt{Ar'^2+By'^2},
\label{eq:entropy-vector-section4}
\end{equation}
where $\dot{\sigma}=H\sigma'$. The sign chosen for $e_s^I$ fixes the
orientation of the entropic direction. Reversing this sign changes the
signs of the entropic perturbation and of the signed turn rate, but
does not affect the physical spectra.

In the spatially flat gauge, the scalar fluctuations are described by
the covariant field perturbations $Q^I$. Expanding the Einstein-frame
action to second order and eliminating the nondynamical lapse and shift
perturbations leads to
\begin{equation}
\mathcal{D}_t^2Q^I
+3H\mathcal{D}_tQ^I
+
\left(
\frac{k^2}{a^2}\delta^I{}_J
+\mathcal{M}^I{}_J
\right)Q^J
=0,
\label{eq:covariant-modes-section4}
\end{equation}
where $\mathcal{D}_t$ is the field-space covariant derivative evaluated
along the background trajectory. The effective mass matrix is
\begin{equation}
\mathcal{M}^I{}_J
=
G^{IK}\nabla_K\nabla_JV
-
R^I{}_{LMJ}\dot{\Phi}^L\dot{\Phi}^M
-
\frac{1}{a^3M_P^2}
\mathcal{D}_t
\left(
\frac{a^3}{H}\dot{\Phi}^I\dot{\Phi}_J
\right),
\label{eq:mass-matrix-section4}
\end{equation}
with $\dot{\Phi}_J=G_{JK}\dot{\Phi}^K$. This expression follows directly
from the covariant quadratic action for multifield perturbations
\cite{Gong:2011uw,Dias:2015rca}. Its three contributions have distinct
origins: the first is the covariant Hessian of the potential, the second
accounts for the intrinsic curvature of field space, and the last
arises from the gravitational constraints. Retaining all three terms
is necessary because neither the background trajectory nor the scalar
manifold is assumed to be effectively single-field.

The perturbations are then projected onto the instantaneous
kinematic basis:
\begin{equation}
Q_\sigma=e_{\sigma I}Q^I,
\qquad
Q_s=e_{sI}Q^I,
\qquad
\mathcal{R}
=
\frac{H}{\dot{\sigma}}Q_\sigma,
\qquad
\mathcal{S}
=
\frac{H}{\dot{\sigma}}Q_s.
\label{eq:RS-definition-section4}
\end{equation}
Here, $\mathcal{R}$ is the comoving curvature perturbation, whereas
$\mathcal{S}$ describes the fluctuation orthogonal to the background
motion. The common normalization makes the two quantities directly
comparable during the numerical evolution.

For each wavenumber, the coupled system is evolved using two
independent mode functions, labeled by $a=1,2$. The curvature and
entropy power spectra are then obtained by projecting these solutions
onto the instantaneous adiabatic and entropy directions:
\begin{align}
{\cal P}_{\cal R}(k)
&=
\frac{k^3}{2\pi^2}
\left(\frac{H}{\dot{\sigma}}\right)^2
\sum_{a=1}^{2}
\left|e_{\sigma I}Q_{(a)}^I\right|^2,
\\
{\cal P}_{\cal S}(k)
&=
\frac{k^3}{2\pi^2}
\left(\frac{H}{\dot{\sigma}}\right)^2
\sum_{a=1}^{2}
\left|e_{sI}Q_{(a)}^I\right|^2.
\label{eq:curvature-entropy-spectra}
\end{align}
These expressions retain the complete coupled evolution and therefore
do not rely on a super-Hubble transfer approximation.

The variation of the tangent vector measures the bending of the
background trajectory. Projecting the background field equation along
the entropic direction gives
\begin{equation}
\eta_\perp
=
e_{sI}\mathcal{D}_Ne_\sigma^I
=
-\frac{V_{,s}}{H\dot{\sigma}},
\qquad
\mathcal{D}_N\equiv H^{-1}\mathcal{D}_t,
\qquad
V_{,s}\equiv e_s^I V_{,I}.
\label{eq:turn-section4}
\end{equation}
Therefore, the trajectory is geodesic when the potential gradient has
no component orthogonal to the motion. A nonzero $\eta_\perp$ couples
the adiabatic and entropic sectors, so that entropy perturbations may
source the curvature perturbation after Hubble crossing.

Projecting the covariant mode equation along $e_s^I$ gives the
effective entropy mass
\begin{equation}
\frac{m_{s,\mathrm{eff}}^2}{H^2}
=
\frac{V_{;ss}}{H^2}
+3\eta_\perp^2
+\epsilon_HM_P^2R_{\mathrm{fs}},
\qquad
V_{;ss}
\equiv
e_s^I e_s^J\nabla_I\nabla_JV.
\label{eq:entropy-mass-section4}
\end{equation}
The first term measures the covariant curvature of the potential in
the direction normal to the trajectory. The second is generated by
the rotation of the kinematic basis, while the last is the geometrical
contribution associated with the curvature of the scalar manifold.
Consequently, the stability of the entropy sector cannot be inferred
from the ordinary potential Hessian alone.

For completeness, the field-space Ricci scalar is obtained by
constructing the Levi--Civita connection of the diagonal metric in
Eq.~\eqref{eq:exact-reduced-metric}. Before substituting the explicit
functions $A(r)$ and $B(r)$, the two-dimensional result can be written
as
\begin{equation*}
R_{\mathrm{fs}}
=
\frac{1}{M_P^2}
\left[
-\frac{B_{,rr}}{AB}
+\frac{A_{,r}B_{,r}}{2A^2B}
+\frac{B_{,r}^2}{2AB^2}
\right].
\end{equation*}
Using the exact radial normalization given previously then yields
\begin{equation}
R_{\mathrm{fs}}
=
\frac{
2\xi_2
\left[
1-4\xi_2^2(1+6\xi_2)r^4
\right]
}{
M_P^2
\left[
1+2\xi_2(1+6\xi_2)r^2
\right]^2
}.
\label{eq:ricci-section4}
\end{equation}
This form is used directly in the numerical evaluation of
Eq.~\eqref{eq:entropy-mass-section4}; no large-field approximation is
introduced.

For each wavenumber, the two independent scalar solutions are
initialized in the covariant Bunch--Davies vacuum at
$k/(aH)=100$. In a locally orthonormal and parallel-transported
field-space basis, labeled by $\hat A=1,2$, the initial conditions are
chosen as
\begin{equation}
Q^{\hat A}_{(a)}
=\frac{\delta^{\hat A}{}_{a}}{a\sqrt{2k}},
\qquad
D_tQ^{\hat A}_{(a)}
=-\left(H+\frac{ik}{a}\right)Q^{\hat A}_{(a)},
\qquad a=1,2.
\label{eq:BD-scalar-initial-conditions}
\end{equation}
These initial conditions are subsequently transformed to the
coordinate basis $(\rho,\phi)$.
At linear order, the tensor perturbations do not mix with the scalar
field-space directions. Each polarization therefore obeys
\begin{equation}
\ddot{h}_{\lambda,k}
+3H\dot{h}_{\lambda,k}
+\frac{k^2}{a^2}h_{\lambda,k}
=0,
\qquad
\lambda=+,\times,
\end{equation}

on the same numerically determined background.
For each polarization, the canonically normalized tensor mode
$v_{\lambda,k}=aM_P h_{\lambda,k}/2$ is initialized in the
Bunch--Davies vacuum according to
\begin{equation}
v_{\lambda,k}
=\frac{e^{-ik\tau}}{\sqrt{2k}},
\qquad
\frac{dv_{\lambda,k}}{d\tau}
=-ikv_{\lambda,k},
\end{equation}
at $k/(aH)=100$.
The total tensor power
is evaluated from both polarizations,
\begin{equation}
\mathcal{P}_T(k)
=
\frac{k^3}{2\pi^2}
\sum_{\lambda=+,\times}
\left|h_{\lambda,k}\right|^2.
\end{equation}
Although the tensor modes evolve independently, the tensor-to-scalar
ratio is affected by the multifield dynamics because the final scalar
spectrum can be enhanced by entropy-to-curvature conversion.

For every wavenumber, two independent scalar modes are initialized in
the covariant Bunch--Davies vacuum at $k/(aH)=100$ and evolved to
$\epsilon_H=1$.  This sub- to super-Hubble evolution follows the
standard numerical strategy used in multifield mode and transport
calculations
\cite{BunchDavies:1978,Price:2014xpa,Ronayne:2017qzn}.  Varying the initial
ratio over $k/(aH)=50,100,200$ changes $n_s$ by less than
$7\times10^{-7}$ and $r_{\rm tens}$ by less than $6\times10^{-9}$ at
the tested scan points.  Changing the finite-difference interval used
for the spectral tilt from $\Delta\ln k=0.015$ to $0.040$ changes
$n_s$ by less than $2\times10^{-8}$.

\begin{figure}[H]
\centering
\includegraphics[width=\linewidth]{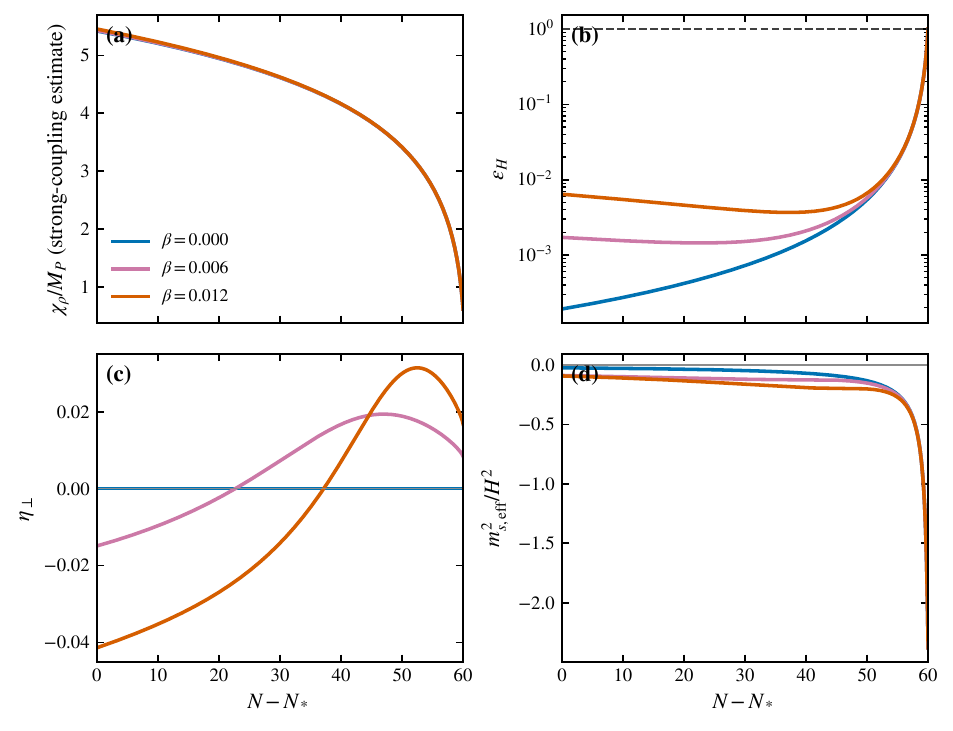}
\caption{Exact background evolution for $N_*=60$ and
$\xi_2=4.9\times10^4$. The numerical solution uses the exact
field-space metric, and inflation ends at $\epsilon_H=1$.}
\label{fig:qidm-background}
\end{figure}

Figure~\ref{fig:qidm-background} illustrates the main
characteristics of the background dynamics for different values of
$\beta$. In panel~(a), the monotonically decreasing trend of
$\chi_\rho/M_P$ reflects the shift of the radial field from the
plateau region towards the end of inflation. The near-superposition of the
curves shows that this radial evolution depends only weakly on
$\beta$. It should be noted, however, that $\chi_\rho$ is presented
here as an estimate in the strong-coupling limit, whilst
the numerical integration is carried out using the exact metric of
the field space.

Panel (b) shows that $\epsilon_H$ remains well below unity
for most of the evolution, which confirms the existence
of a prolonged quasi-de Sitter phase. A larger value of
$\beta$ initially increases $\epsilon_H$ due to the kinetic
contribution from the quintessence component. Nevertheless, all
trajectories converge towards $\epsilon_H=1$, which marks the end of
inflation.

Panel (c) highlights a signed turn rate
$\eta_\perp$ that is small but non-zero when $\beta\neq0$. Its magnitude
increases with $\beta$, whilst its change in sign reflects a
change in the orientation of the trajectory’s curvature in
field space. Although the instantaneous rotation remains small, its
integrated effect over several tens of e-folds can produce a
cumulative transfer of entropic perturbations towards the
curvature perturbation.

Finally, panel (d) shows that the effective entropic mass
$m_{s,\mathrm{eff}}^2/H^2$ remains negative and of small magnitude
for most of inflation, before decreasing rapidly as
it nears its end. This behaviour corresponds to a
slightly tachyonic entropic direction, capable of amplifying
entropy perturbations without causing immediate destabilisation
of the background trajectory. The combined action of this negative mass and
the nonzero turn rate thus provides the physical mechanism allowing
a significant entropy-curvature transfer despite the small
instantaneous value of $\eta_\perp$.

\subsection{CMB observables and covariant de Sitter diagnostics}
\label{subsec:CMB-observables}
\label{subsec:swampland-results}

The curvature and entropy spectra defined above are evaluated at the
end of inflation. At the pivot scale
$k_*=0.05\,\mathrm{Mpc}^{-1}$, we impose
\begin{equation}
{\cal P}_{\cal R}^{\rm end}(k_*)=A_s^{\rm obs}
=2.10\times10^{-9}.
\label{eq:As-normalization-section4}
\end{equation}
At fixed $(\xi_2,\beta,N_*,y_*)$, the dimensionless trajectory is
independent of the overall quartic normalization. A reference run with
$\lambda_2^{\rm ref}=1$ therefore gives
\begin{equation}
\lambda_2^{\rm norm}
=
\frac{A_s^{\rm obs}}
{{\cal P}_{\cal R}^{{\rm end},{\rm ref}}(k_*)}.
\label{eq:lambda-normalization-section4}
\end{equation}
Equation~\eqref{eq:lambda-normalization-section4} fixes the
normalization governed by $\lambda_2/\xi_2^2$ independently at each
point of the scan. It does not select $\beta\simeq0.004$ or any other
preferred value of the exponential slope.

The scalar spectral index and tensor-to-scalar ratio are evaluated as
\begin{equation}
n_s-1
=
\left.
\frac{d\ln{\cal P}_{\cal R}^{\rm end}}
{d\ln k}
\right|_{k_*},
\qquad
r_{\rm tens}
=
\frac{{\cal P}_T^{\rm end}(k_*)}
{{\cal P}_{\cal R}^{\rm end}(k_*)}.
\label{eq:ns-r-section4}
\end{equation}
The notation $r_{\rm tens}$ distinguishes the tensor-to-scalar ratio
from the radial coordinate $r$. We also introduce the
end-of-inflation entropy fraction
\begin{equation}
\beta_{\rm iso}^{\rm end}
=
\frac{{\cal P}_{\cal S}^{\rm end}}
{{\cal P}_{\cal R}^{\rm end}
+{\cal P}_{\cal S}^{\rm end}}.
\label{eq:beta-iso-end}
\end{equation}
This quantity is an inflationary endpoint diagnostic and should not be
identified directly with a CMB isocurvature fraction without specifying
the reheating process and the subsequent evolution of the entropy mode.

\begin{figure}[H]
\centering
\includegraphics[width=\linewidth]{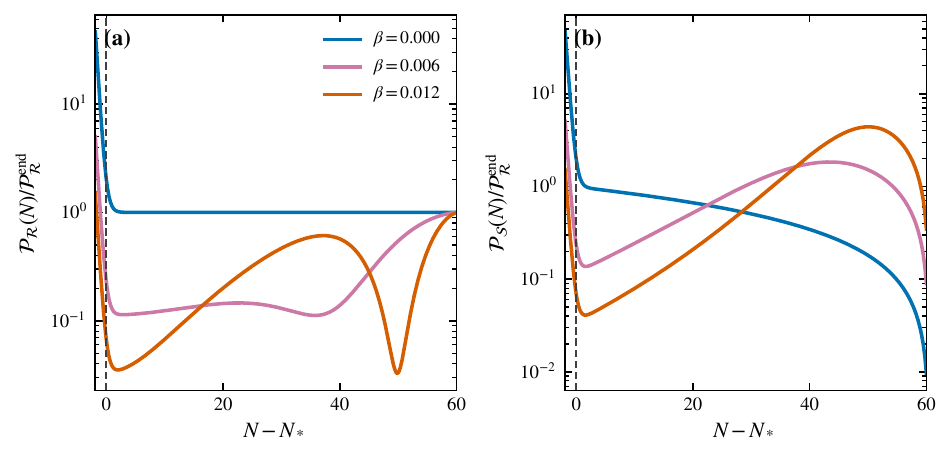}
\caption{Evolution of the normalized curvature and entropy spectra at
the pivot scale for $N_*=60$. The dashed line marks Hubble crossing,
and both independent scalar modes are included.}
\label{fig:qidm-modes}
\end{figure}

Figure~\ref{fig:qidm-modes} illustrates the distinct roles of the
adiabatic and entropic sectors during the super-Hubble evolution. In
panel~(a), the curvature spectrum for $\beta=0$ rapidly approaches a
constant value after Hubble crossing, as expected for an approximately
geodesic trajectory with negligible mode coupling. For $\beta>0$,
however, the curvature spectrum exhibits a prolonged and non-monotonic
evolution. Since this variation occurs when the mode is already outside
the Hubble radius, it cannot be attributed to spatial-gradient effects.
It instead signals the continuous sourcing of the curvature
perturbation by the entropy mode through the nonvanishing turn rate of
the background trajectory. The increasing departure from conservation
with larger $\beta$ reflects a stronger cumulative multifield transfer.

Panel~(b) clarifies the origin of this behaviour. In the decoupled
limit, the entropy spectrum decreases progressively, whereas for
$\beta>0$ it undergoes a substantial intermediate amplification. This
growth is consistent with the slightly negative effective entropy mass
found along the inflationary trajectory. The subsequent evolution of
the entropy spectrum is accompanied by a modification of the curvature
spectrum, providing direct evidence for entropy-to-curvature
conversion. Because each curve is normalized by the final curvature
power of the corresponding parameter point, the figure displays the
relative time evolution rather than a comparison of absolute
amplitudes between different values of $\beta$. These results show that
the observable scalar spectrum must be evaluated after the multifield
transfer has taken place; consequently, the usual single-field
horizon-exit relations, including $r_{\rm tens}=16\epsilon_{H*}$, are
not applicable to the present dynamics
\cite{Gordon:2000hv,Wands:2002bn,Achucarro:2012sm}.

Representative results for $N_*=60$ are collected in
Table~\ref{tab:qidm-scan}. The points $\beta=0.006$ and $0.012$ are
included to illustrate the evolution across the sampled parameter
range; no special role is assigned to $\beta=0.004$.

\begin{table}[H]
\centering
\caption{Representative multifield results for
$\xi_2=4.9\times10^4$ and $N_*=60$.}
\label{tab:qidm-scan}
\begin{tabular}{c c c c c c}
\hline\hline
$\beta$ & $\lambda_2$ & $n_s$ & $r_{\rm tens}$
& $\int|\eta_\perp|\,dN$ & $\beta_{\rm iso}^{\rm end}$\\
\hline
$0$     & $0.9195$ & $0.96775$ & $0.00301$ & $0$     & $0.0096$\\
$0.006$ & $0.9537$ & $0.96560$ & $0.00310$ & $0.675$ & $0.0823$\\
$0.012$ & $1.0530$ & $0.95894$ & $0.00340$ & $1.447$ & $0.2578$\\
\hline
\end{tabular}
\end{table}

\begin{figure}[H]
\centering
\includegraphics[width=\linewidth]
{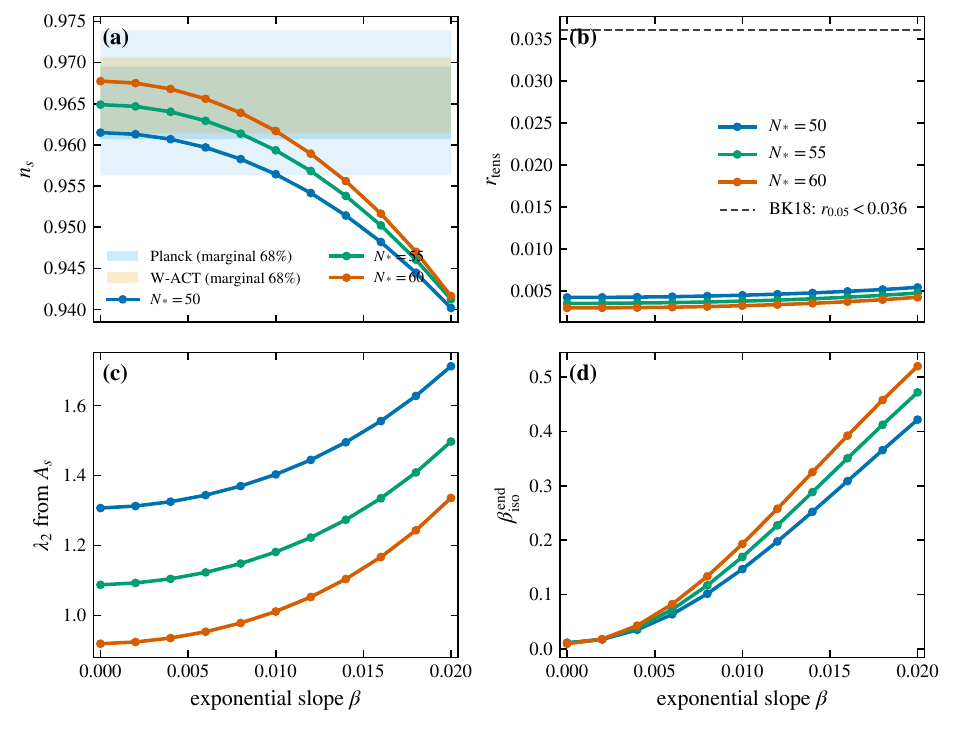}
\caption{Final inflationary observables and amplitude normalization as
functions of $\beta$. The shaded regions indicate marginalized CMB
intervals, while panel~(d) shows the entropy fraction before reheating}
\label{fig:qidm-cmb-scan}
\end{figure}

The complete scan is displayed in
Fig.~\ref{fig:qidm-cmb-scan}. Throughout the sampled domain,
$r_{\rm tens}$ lies between $3.0\times10^{-3}$ and
$5.5\times10^{-3}$ and therefore remains well below the BICEP/Keck
BK18 limit $r_{0.05}<0.036$ at $95\%$ confidence
\cite{Ade:2021rcr}. The scalar spectral index provides a stronger
sensitivity to $\beta$. For comparison, the ACT DR6 analysis gives
$n_s=0.9660\pm0.0046$ for W-ACT and
$n_s=0.9651\pm0.0044$ for its Planck comparison
\cite{Louis:2025x}.

Interpolating the present scan against the lower boundary of the
marginal two-standard-deviation Planck interval gives the indicative
limits
\begin{equation}
\beta\lesssim0.010,\ 0.012,\ 0.014
\qquad\text{for}\qquad
N_*=50,\ 55,\ 60,
\label{eq:indicative-beta-limits}
\end{equation}
respectively. The corresponding W-ACT estimates are approximately
$0.010$, $0.012$, and $0.013$. These values are sensitivity estimates
derived from one-dimensional marginalized intervals and do not replace
a joint likelihood analysis in
$(n_s,r_{\rm tens},\beta_{\rm iso})$.

The substantial increase of $\beta_{\rm iso}^{\rm end}$ at larger
$\beta$ is as relevant as the corresponding variation of $n_s$.
Consequently, a complete observational constraint requires an explicit
reheating prescription and a treatment of the subsequent entropy-mode
evolution. Until such an analysis is performed,
Eq.~\eqref{eq:indicative-beta-limits} should be interpreted as a
conditional sensitivity estimate rather than as a definitive CMB
bound.

Having characterized the observational behaviour of the model, we
evaluate the covariant de Sitter diagnostics along the same numerical
trajectories. This allows the phenomenologically relevant region of
the parameter scan to be compared directly with the gradient and
Hessian branches of the refined de Sitter conjecture.

The relevant gradient diagnostic during inflation is the complete
field-space norm defined in Eq.~\eqref{eq:ceff-definition}.
Substitution of the exact reduced metric and potential gives
\begin{equation}
c_{\rm eff}^2
=
\frac{16}
{r^2\left(2F+24\xi_2^2r^2\right)}
+\beta^2F.
\label{eq:ceff-exact-section4}
\end{equation}
The contribution $2F$ in the first denominator follows from the
corrected radial normalization. More importantly,
Eq.~\eqref{eq:ceff-exact-section4} shows explicitly that
$c_{\rm eff}$ cannot be identified with the exponential slope
$\beta$ during the inflationary multifield evolution.

\begin{figure}[H]
\centering
\includegraphics[width=\linewidth]
{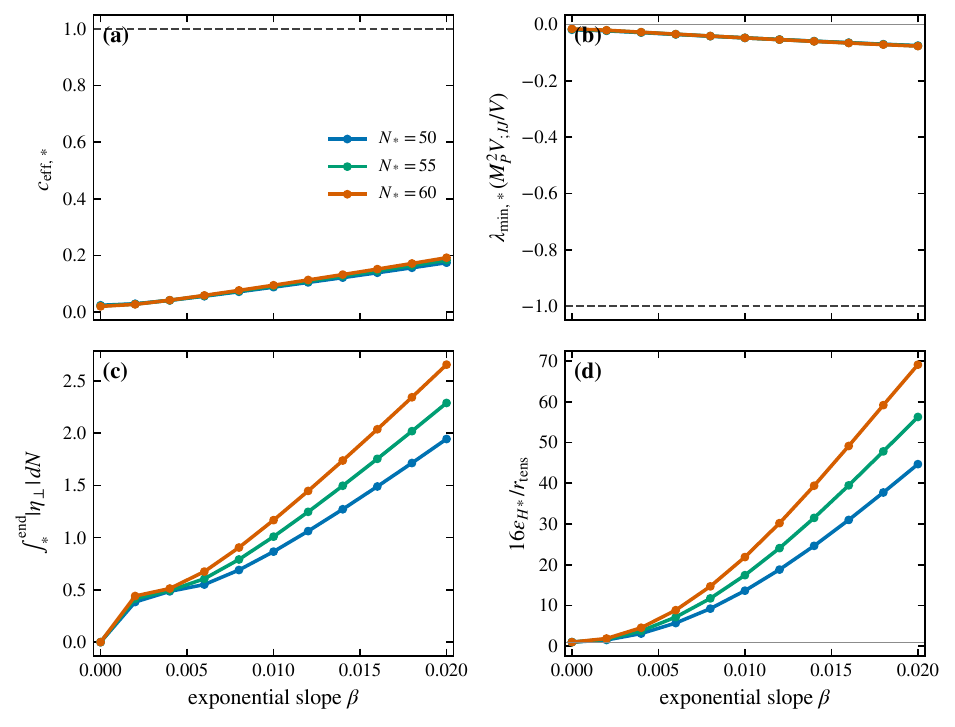}
\caption{Covariant inflationary diagnostics across the parameter scan.
The dashed lines mark the order-unity Swampland reference values.}
\label{fig:qidm-covariant-scan}
\end{figure}

For the refined Hessian branch, we evaluate the smallest eigenvalue of
the mixed covariant matrix defined in
Eq.~\eqref{eq:normalized-covariant-Hessian}.
Figure~\ref{fig:qidm-covariant-scan} displays the gradient and Hessian
diagnostics together with the integrated turn and the scalar
enhancement measure $16\epsilon_{H*}/r_{\rm tens}$.

Within the portion of the scan compatible with the indicative
two-standard-deviation scalar-tilt intervals, we find
$c_{\rm eff,*}\lesssim0.13$ and
$\lambda_{\min,*}\gtrsim-0.06$. Thus, neither the gradient diagnostic
nor the magnitude of the smallest Hessian eigenvalue becomes
order unity during the CMB-relevant stage of inflation. The
multifield transfer substantially affects the observable scalar
spectrum, but it does not remove the tension with a strong
order-unity interpretation of the refined de Sitter conjecture.

\subsection{Late-time evolution and consistency checks}
\label{subsec:late-time-evolution}
\label{subsec:IDM-stability-results}

After stabilization of the radial direction, the residual scalar
potential is taken to be
\begin{equation}
V_{\rm DE}(\phi)
=
\Lambda_{\rm eff}e^{-\beta\phi/M_P}.
\label{eq:late-potential-section4}
\end{equation}
To isolate the late-time consequences of the slopes selected by the
inflationary scan, we perform an effective post-reheating calculation
using the standard phase-space formulation of exponential
quintessence
\cite{FerreiraJoyce:1997,RatraPeebles:1987}. We introduce
\begin{equation}
x_q=\frac{\dot\phi}{\sqrt6HM_P},
\qquad
y_q=\frac{\sqrt{V_{\rm DE}}}{\sqrt3HM_P},
\qquad
\Omega_\phi=x_q^2+y_q^2.
\end{equation}
Here and throughout the late-time analysis, a prime denotes
differentiation with respect to the number of e-folds,
$N=\ln a$. We normalize the present scale factor to $a_0=1$ and
define the cosmological redshift by
\begin{equation}
1+z=\frac{a_0}{a}=\frac{1}{a},
\qquad
N=-\ln(1+z),
\qquad
\frac{d}{dN}=-(1+z)\frac{d}{dz}.
\label{eq:redshift-efold-relation}
\end{equation}
The matter and radiation density parameters are defined by
\begin{equation}
\Omega_m=\frac{\rho_m}{3M_P^2H^2},
\qquad
\Omega_r=\frac{\rho_r}{3M_P^2H^2}.
\label{eq:matter-radiation-fractions}
\end{equation}
The Klein--Gordon equation for $\phi$, together with the matter and
radiation continuity equations, gives
\begin{equation}
\frac{H'}{H}
=-\frac{3}{2}
\left(
2x_q^2+\Omega_m+\frac{4}{3}\Omega_r
\right).
\label{eq:late-Hubble-derivative}
\end{equation}
Using this relation and including radiation and pressureless matter,
the late-time evolution is governed by the autonomous system
\begin{align}
x_q'
&=
-3x_q+\sqrt{\frac32}\beta y_q^2
+\frac32x_q
\left(
2x_q^2+\Omega_m+\frac43\Omega_r
\right),
\\
y_q'
&=
-\sqrt{\frac32}\beta x_qy_q
+\frac32y_q
\left(
2x_q^2+\Omega_m+\frac43\Omega_r
\right),
\\
\Omega_r'
&=
\Omega_r
\left[
-4+3
\left(
2x_q^2+\Omega_m+\frac43\Omega_r
\right)
\right],
\end{align}
where
\begin{equation}
\Omega_m=1-x_q^2-y_q^2-\Omega_r,
\qquad
w_\phi
=
\frac{x_q^2-y_q^2}{x_q^2+y_q^2}.
\end{equation}

We initialize the late-time integration at
$z_i=10^7$, deep in the radiation-dominated epoch, and impose
$x_q(z_i)=0$, corresponding to an initially frozen quintessence
field. The normalization $\Lambda_{\rm eff}$ is then adjusted so that
$\Omega_{\phi0}=0.685$ at $z=0$.

\begin{figure}[H]
\centering
\includegraphics[width=\linewidth]
{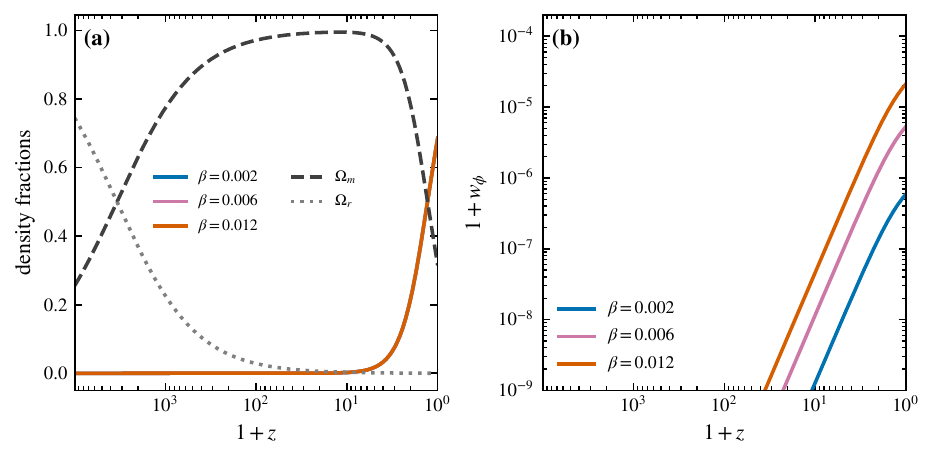}
\caption{Effective late-time quintessence evolution normalized to
$\Omega_{\phi0}=0.685$.}
\label{fig:qidm-late-time}
\end{figure}

These results show that the small exponential slopes compatible with
the inflationary analysis produce a background evolution extremely
close to $\Lambda$CDM once the residual dark-energy scale is fixed.
This effective treatment does not, however, explain the hierarchy in
$\Lambda_{\rm eff}$ or determine the initial post-reheating scalar
velocity. Both aspects require an explicit reheating and dynamical
matching prescription.

The validity of this reduced cosmological description also requires
the scalar directions orthogonal to the retained two-field trajectory
to remain sufficiently stabilized. We therefore examine the leading
masses of the neglected IDM fluctuations along the inflationary
background.

The configuration
\begin{equation*}
H_2=(0,\rho)^T
\end{equation*}
selects a radial direction in the inert doublet. The remaining
components of $H_2$ correspond to angular gauge directions during this
phase and should not be identified directly with three independent
physical scalars carrying their electroweak-vacuum masses. The relevant
scalar stability test instead concerns the four real components of the
orthogonal active doublet $H_1$.

For $\xi_1=0$, their leading large-field potential masses are governed
by
\begin{equation}
c_h
=
\lambda_3+\lambda_4+\lambda_5
=
2\lambda_L,
\qquad
c_a
=
\lambda_3+\lambda_4-\lambda_5,
\qquad
c_\pm
=
\lambda_3.
\label{eq:orthogonal-couplings}
\end{equation}
Up to covariant connection and field-space-curvature contributions,
which are of order $H^2$ in the regime considered, the corresponding
mass ratios are
\begin{equation}
\frac{m_{X,\perp}^2}{H^2}
\simeq
\frac{3c_XF}{\lambda_2r^2},
\qquad
X=h,a,\pm.
\label{eq:orthogonal-mass-estimate}
\end{equation}

As a representative tree-level benchmark, we take
\begin{equation}
m_S=600~{\rm GeV},
\qquad
m_A=605~{\rm GeV},
\qquad
m_{H^\pm}=610~{\rm GeV},
\qquad
\lambda_L=0.005.
\end{equation}
The electroweak relations then give approximately
\begin{equation}
\lambda_3=0.410,
\qquad
\lambda_4=-0.300,
\qquad
\lambda_5=-0.100,
\end{equation}
so that all three coefficients in
Eq.~\eqref{eq:orthogonal-couplings} are positive. Over the
$N_*=60$ scan points satisfying $0\leq\beta\leq0.014$, the smallest
leading mass ratios are
\begin{equation}
\frac{m_{h,\perp}^2}{H^2}
\gtrsim2.7\times10^3,
\qquad
\frac{m_{a,\perp}^2}{H^2}
\gtrsim5.6\times10^4,
\qquad
\frac{m_{\pm,\perp}^2}{H^2}
\gtrsim1.1\times10^5.
\label{eq:orthogonal-mass-results}
\end{equation}
These large positive ratios show that the orthogonal active-doublet
fluctuations are heavy compared with the Hubble scale. The reduced
two-field trajectory is therefore stable against these tree-level
fluctuations for the representative benchmark considered.

This test should nevertheless be interpreted as a consistency check
of the reduced inflationary trajectory rather than as a complete
high-scale analysis of the IDM. A nonzero $\xi_1$, the renormalization
group evolution of the quartic and portal couplings, gauge
fluctuations, loop corrections, and matching to relic-density and
collider constraints must be incorporated before claiming a fully
ultraviolet-stable IDM benchmark. Keeping these limitations explicit
separates the robust cosmological consequences of the two-field
dynamics from the model-dependent particle-physics assumptions.

\section{Conclusion}
\label{sec:conclusion}

In this work, we investigated inflation in the Inert Doublet Model
coupled to a light quintessence field through an exponential
interaction. In the Einstein frame, the inflationary dynamics is
described by a noncanonical two-field system with a curved field-space
geometry. We treated both the background evolution and the primordial
perturbations within a fully covariant multifield formalism.

Our numerical analysis shows that the quintessence direction affects
the inflationary predictions through the cumulative bending of the
background trajectory. Although the instantaneous turn rate remains small but nonzero, the associated entropy-to-curvature transfer modifies the
final scalar spectrum and demonstrates that horizon-crossing
single-field estimates are insufficient. The scalar amplitude fixes
the overall inflationary normalization for each trajectory, while the
exponential slope must be constrained independently through a scan of
the multifield observables. Within the explored parameter space, the
scalar spectral index provides the strongest sensitivity, whereas the
tensor-to-scalar ratio remains compatible with current observational
limits.

The covariant Swampland diagnostics indicate that multifield effects
do not remove the tension with a strong order-unity interpretation of
the refined de Sitter conjecture during the observable stage of
inflation. At late times, the residual exponential sector can produce
an expansion history close to that of the standard cosmological model
once its energy scale is normalized to the observed dark-energy
density. The reduced inflationary trajectory is also stable against
the leading orthogonal scalar fluctuations for the representative IDM
configuration considered.

A definitive assessment of the scenario requires a dynamical
description of reheating and of the subsequent evolution of the
entropy perturbations. The running of the scalar couplings, quantum
corrections, and the connection with dark-matter and collider
constraints should also be incorporated into a complete
phenomenological analysis. These extensions will determine whether the
model can simultaneously provide a consistent description of
inflation, dark energy, and the inert scalar sector.
\section*{Acknowledgments}
The work of S. Saoud, M. A. Rbah and R. Sammani  is funded by the National Center for Scientific and Technical
Research (CNRST) under the PhD-Associate Scholarship (PASS).
\bibliographystyle{unsrt}
\bibliography{Bib}

\end{document}